\documentclass[a4paper,10pt,twoside]{cpc-hepnp}

\usepackage{multicol}
\usepackage{graphicx}
\usepackage{booktabs}
\usepackage{amssymb,bm,mathrsfs,bbm,amscd}
\usepackage[tbtags]{amsmath}
\usepackage{lastpage}
\usepackage{CJK}
\usepackage{float}

\def\bea{\begin{eqnarray}}
\def\eea{\end{eqnarray}}
\def\beas{\begin{eqnarray*}}
\def\eeas{\end{eqnarray*}}
\def\be{\begin{equation}}
\def\ee{\end{equation}}
\def\bes{\begin{equation*}}
\def\ees{\end{equation*}}
\def\onefigure{\includegraphics}
\def\lsim{\raise0.3ex\hbox{$\;<$\kern-0.75em\raise-1.1ex\hbox{$\sim\;$}}}
\def\gsim{\raise0.3ex\hbox{$\;>$\kern-0.75em\raise-1.1ex\hbox{$\sim\;$}}}

\def\nn{\nonumber}

\newcommand{\lm}{\lambda}

\newcommand{\al}{\alpha}

\begin{document}
%\begin{CJK*}{GBK}{song}

\fancyhead[c]{\small Chinese Physics C~~~Vol. 42, No. 9 (2018)
093107} \fancyfoot[C]{\small 093017-\thepage}

\footnotetext[0]{Received 23 April 2018}

\title{$HZ$ associated production with decay in the Alternative Left-Right Model at CEPC and future linear colliders\thanks{Supported by the National Natural Science Foundation of China (No.11375008, No.11647307) }}

\author{%
     Juan-Juan Niu$^{1}$%
\quad Lei Guo $^{1;1)}$\email{guoleicqu@cqu.edu.cn, Corresponding author}%
\quad Shao-Ming Wang $^{1}$%
}

\maketitle

\address{%
$^1$ Department of Physics,~Chongqing University,~Chongqing 401331,~P.R. China and Institute of Theoretical Physics,~Chongqing University,~Chongqing 401331,~P.R. China.
}

\begin{abstract}
In this study, Higgs and Z boson associated production with subsequent decay is attempted in the framework of
alternative left-right model, which is motivated by superstring-inspired $E_6$ model at CEPC and future linear colliders. We systematically analyze each
decay channel of Higgs with theoretical constraints and latest experimental methods.
Due to the mixing of scalars in the Higgs sector, charged Higgs bosons can play an essential role in the phenomenological analysis of this process.
Even though the predictions of this model for the signal strengths of
this process are close to the standard model expectations, it can be distinct under high luminosity.
\end{abstract}



\footnotetext[0]{\hspace*{-3mm}\raisebox{0.3ex}{$\scriptstyle\copyright$}2013
Chinese Physical Society and the Institute of High Energy Physics
of the Chinese Academy of Sciences and the Institute
of Modern Physics of the Chinese Academy of Sciences and IOP Publishing Ltd}%

\begin{multicols}{2}

\section{Introduction }

With the discovery of Higgs boson by ATLAS~\cite{Aad:2012tfa} and CMS~\cite{Chatrchyan:2012xdj}
at Large Hadron Collider(LHC) in~2012, the standard model(SM)
has made a great accomplishment. Higgs production with subsequent decay plays an essential role
not only in the precision test of the Higgs property but also provides a window to new physics beyond the SM(BSM).
The study of Higgs and Z boson associated production and decay at Higgs factory such as Circular Electron
Positron Collider(CEPC) and future linear colliders
is significantly important in measuring gauge and Yukawa interactions, so more and more
theorists and experimenters are motivated to investigate this process in new physics scenarios~\cite{Thomson:2015jda,Greco:2017fkb,Mahanta:1997wd,Craig:2014una}.
CEPC was proposed by Chinese scientists at about 240 GeV center-of-mass energy mainly for Higgs studies
with two detectors situated in a very long tunnel more than twice the size of the LHC at CERN.
A future linear collider, such as International Linear Collider(ILC) or Compact Linear Collider(CLIC), at center-of-mass energy
$\sqrt{s}=500$ GeV or even higher in the TeV energy scale, will allow the Higgs sector to be probed
with high precision significantly beyond that at High-Luminosity
LHC~\cite{CEPC-SPPCStudyGroup:2015csa,Baer:2013cma,Abramowicz:2013tzc}.
CEPC and future linear colliders are $e^+~e^-$ colliders and
will be crucial facilities for precision Higgs physics research, outcome of which may be an order of magnitude more precise than that achievable
at LHC. Such measurements may be necessary to reveal BSM effects in Higgs sector.
Moreover, $e^+~e^-$ colliders provide an opportunity to measure Higgs couplings, rather than ratios, with a cleaner background. In addition, an $e^+~e^-$ collider operating at 1 TeV
or above, for example CLIC or an upgraded ILC, will have the sensitivity to top quark Yukawa coupling and Higgs self-coupling
parameters, and thus will provide a direct probe of Higgs potential.

As for the discovery of neutrino masses and neutrino oscillations, it confirmed that
SM remains incomplete. To provide a proper explanation for the measured
neutrino masses, theorists have made several attempts,
such as supersymmetry, extra dimensions, Two Higgs Doublet Model(2HDM), and Left-Right Model (LRM), to expand the SM. The Alternative Left-Right
Model(ALRM)~\cite{Ma:1986we,Babu:1987kp,1742-6596-315-1-012006,Ma:2012gb}, motivated
by the superstring-inspired $E_6$ model, is a type of left-right
model~\cite{Mohapatra:1974gc,Senjanovic:1975rk,Maiezza:2010ic,Borah:2010zq}.
ALRM is based on $SU(3)_C \times SU(2)_L \times
SU(2)_R \times U(1)_{(B-L)/2} \times S$, where $SU(2)_R \times U(1)_{(B-L)/2}$
can break at TeV scale, allowing several interesting signatures
at LHC. In ALRM all non-SM particles can couple
with SM fermions and Higgs bosons which will lead to low energy consequences. Due to the
rich Higgs sector, there are four neutral CP-even and two CP-odd Higgs
bosons, in addition to two charged Higgs bosons, which come from one bidoublet and two left-handed
and right-handed doublets; most of these Higgs bosons can be light, falling in the
electroweak scale. As for the couplings of the SM-like Higgs with the fermions and
gauge bosons, there are small changes compared to the corresponding ones in SM. In the
literature, many studies have been undertaken, which primarily focus on dark matter or hadron production
of ALRM~\cite{1742-6596-315-1-012006,Khalil:2010yt,Ashry:2013loa,Mandal:2017tab}.

In this study, we will mainly focus on the weak production of Z boson and Higgs.
Each channel of Higgs decay modes has been analyzed in ALRM and
compared with the recent results reported by ATLAS and CMS experiments~\cite{Sirunyan:2017exp,Chatrchyan:2013zna,Aaboud:2017vzb,Aaboud:2017xsd,Khachatryan:2016vau}.
There are possible discrepancies between the
results of signal decay strengths in each channel. We analyze through five main Higgs decay channels:
$H \to b\bar{b},~c\bar{c},~\tau\bar{\tau},~ZZ$ and $WW$ in both ALRM and SM correspondingly.
We find that the signal strength of Higgs decay channel
$H \to ZZ$ is consistent with SM expectation. However, the decay channel $H \to b\bar{b}$
is more sensitive to the mass of charged Higgs, where there may exist discrepancy
with SM. Secondly, $HZ$ associated production has been
systematically explored in this model at CEPC and future $e^+~e^-$ colliders.
The couplings of new heavy bosons to the known fundamental particles
will be a crucial test of SM and may support an opportunity to establish physics
BSM. We find that the discrepancies of cross-sections between
ALRM and SM are of a few percent at $\sqrt{s}=240$ GeV, $500$ GeV, and
1 TeV. Finally, we study the subsequent decay of the final state Higgs into a pair of
bottom quarks and Z boson into a leptonic pair, where $l = e,~\mu$ and $\tau$,
associated with ALRM and the required integrated luminosity when the discovery significance is $5\sigma$.

The organization of this paper is as follows: In section 2 we briefly describe
the related theory of ALRM. In section 3, we perform the numerical analysis for
Higgs and Z boson associated production with decay in ALRM.
Finally, a short summary is provided in section 4.

\section{Alternative Left-Right Symmetric Model}

ALRM is a standard model extension based on $SU(3)_C \times SU(2)_L \times
SU(2)_R \times U(1)_{(B-L)/2} \times S$ gauge symmetry, the discrete
symmetry $S$ is to distinguish the scalar bidoublet from its dual
scalars.
The details of ALRM are reported in the literature~\cite{Ashry:2013loa}.
Here, we only introduce the formulas used in our calculations.

Let us start from the most general left-right symmetric Yukawa Lagrangian:

\bea
\mathcal{L}_\text{Y}&=&
\overline{Q}_LY^q\widetilde{\Phi}Q_R+\overline{Q}_L Y^q_L\chi_L d_R+\overline{Q}_R Y^q_R\chi_R d'_L\nn
\\&+&\overline{\psi}_L Y^\ell\Phi\psi_R\nn
+\overline{\psi}_L Y^\ell_L\widetilde{\chi}_L\nu_R+\overline{\psi}_R Y^\ell_R\widetilde{\chi}_R n_L\nn
\\&+&\overline{\nu}^c_R M_R\nu_R + \text{h.c.}\,
\label{Yukawa}
\eea
In Eq.(\ref{Yukawa}), $\widetilde{\Phi}$ and $\widetilde{\chi}_{L,R}$ are the duals of the bidoublet
$\Phi$ and the doublets $\chi_{L,R}$, which are defined as $\widetilde{\Phi}=\tau_2\Phi^*\tau_2$
and $\widetilde{\chi}_{L,R}=i\tau_2\chi^*_{L,R}$. From this Lagrangian,
we can get the masses of the fermions, which are quarks $u$, $d$, $d'$, the
charged leptons $\ell$, and the additional singlet fermion $n$ called scotino.
\bea
&&m_u=\frac{1}{\sqrt{2}}Y^q v\sin\beta,~~~~
m_d=\frac{1}{\sqrt{2}}Y^q_L v\cos\beta,\nn\\
&&m_{d'}=\frac{1}{\sqrt{2}}Y^q_R v_R,~~~~~~~~
m_\ell=\frac{1}{\sqrt{2}}Y^\ell v\sin\beta,\nn\\
&&m_{n}=\frac{1}{\sqrt{2}}Y^\ell_R v_R.
\eea
The mixing angle and the vacuum expectation are set as $\tan\beta=k/v_L$ and $\sqrt{v_L^2 + k^2} = v \equiv 246 ~ {\rm GeV}$.
From Eq.(\ref{Yukawa}), we also get the Yukawa couplings of the SM-like Higgs
\bea
&&Y_{H\bar{u}u}=\frac{m_u}{v}\frac{T_{\Phi}}{\sin\beta},~~~~
Y_{H\bar{d}d}=\frac{m_d}{v}\frac{T_{L}}{\cos\beta},\nn\\
&&Y_{H\bar{d}'d'}=\frac{m_{d'}}{v_R}T_{R},~~~~~~
Y_{H\bar{l}l}=\frac{m_l}{v}\frac{T_{\Phi}}{\sin\beta},\nn\\
&&Y_{H\bar{n}n}=\frac{m_{n}}{v_R}T_{R},\label{Qcoup}
\label{Fcoup}
\eea
where the $T_{\Phi},\ T_{L}$ and $T_{R}$ are the mixing parameters
of the SM-like Higgs with the gauge eigenstates $\phi_2^{0R},\ \chi_L^{0R}$ and $\chi_R^{0R}$, respectively.
Similar to the Yukawa coupling, the specific couplings of the SM-like Higgs with the massive EW gauge
bosons can also be derived from the Lagrangian of the scale sector~\cite{Ashry:2013loa}.

In the gauge sector,
$W^\pm_L$ and $W^\mp_R$ cannot mix with each other as $<\phi_1^0>=0$.
We get the mass eigenstates $W^\pm=W_L^\pm$, which are the SM gauge bosons,
and the heavy charged bosons $W'^\pm=W_R^\pm$. The masses of these bosons are given by:
\bea
&&M_W^2=\frac{1}{4}g^2\left(k^2+v_L^2\right)=\frac{1}{4}g^2 v^2,\\
&&M_{W'}^2=\frac{1}{4}g^2\left( k^2+v_R^2\right).
\eea
At present, numerous measurements focused on the heavy bosons have been conducted.
Till date, a search for high-mass resonances using an integrated luminosity of 36.1 $fb^{-1}$ by the ATLAS collaboration offers a lower mass limit of $W'$ as
$M_{W'} \gsim 3.7 $ TeV~\cite{Aaboud:2018vgh,Aaboud:2016zkn,Olive:2016xmw}. This lower boundary on $M_{W'}$ is
applicable in our following calculation in this study.

For the neutral gauge bosons, the masses of two massive bosons can be calculated using
\bea
&&M_{Z,Z'}^2=\frac{1}{2}(M_{LL}^2+M_{RR}^2\mp\left(M_{RR}^2-M_{LL}^2\right)\times\nn\\
&&~~~~~~~~~~~~~\sqrt{1+\tan^22\vartheta}),
\label{Z-Z' Mass}
\eea
where mixing angle $\vartheta$ is defined as
\be
\tan2\vartheta=\frac{2M_{LR}^2}{M_{LL}^2-M_{RR}^2},\label{Z-Z' Mixing}
\ee
and
\begin{equation}
\hspace{-0.17in}
\begin{array}{l}
\displaystyle M_{LL}^2=\frac{g^2 v^2}{4\cos^2\theta_w},\\
\displaystyle M_{LR}^2=\frac{g^2 (v^2\sin^2\theta_w-k^2 \cos^2\theta_w)}{4\cos^2\theta_w\sqrt{\cos2\theta_w}},\\
\displaystyle M_{RR}^2=\frac{g^2 (2v^2\sin^4\theta_w+2(k^2+v_R^2)\cos^4\theta_w-k^2\sin^22\theta_w)}{8\cos^2\theta_w \cos2\theta_w}.\nonumber
\end{array}
\end{equation}
From Eqs.~(\ref{Z-Z' Mass}-\ref{Z-Z' Mixing}), we know that the mixing angle $\vartheta$ strongly influences the masses of $Z$ and $Z'$, and when $\vartheta \to 0$,
$Z\simeq Z_L$ and $Z' \simeq Z_R$. The latest LHC experiments using a data sample corresponding to an integrated luminosity of $36.1 fb^{-1}$ from proton-proton collisions give a limitation for the $Z'$ gauge
boson as $M_{Z'} \gsim 2.42$ TeV~\cite{Aaboud:2017sjh,Aaboud:2016cth,Olive:2016xmw}, and we use the constraint in our numerical calculation.

Reference~\cite{Borah:2010zq} gives the most general Higgs potential with the symmetry invariance,
\bea
&&V(\Phi,\chi_{L,R})~=~
-\mu_1^2Tr[\Phi^\dag\Phi]+\lm_1(Tr[\Phi^\dag\Phi])^2\nn\\
&&~~~~~~+\lm_2 Tr[\Phi^\dag\widetilde{\Phi}] ~ Tr[\widetilde{\Phi}^\dag\Phi] -\mu_2^2(\chi_L^\dag\chi_L+\chi_R^\dag\chi_R)\nn\\
&&~~~~~~+\lm_3[(\chi_L^\dag\chi_L)^2+(\chi_R^\dag\chi_R)^2]+2\lm_4(\chi_L^\dag\chi_L)(\chi_R^\dag\chi_R)\nn
\\&&~~~~~~+2\al_1Tr(\Phi^\dag\Phi)(\chi_L^\dag\chi_L+\chi_R^\dag\chi_R)+2\al_2(\chi_L^\dag\Phi\Phi^\dag\chi_L\nn\\
&&~~~~~~+\chi_R^\dag\Phi^\dag\Phi\chi_R)+2\al_3(\chi_L^\dag\widetilde{\Phi}\widetilde{\Phi}^\dag\chi_L+\chi_R^\dag\widetilde{\Phi}^\dag\widetilde{\Phi}\chi_R)
\nn\\&&~~~~~~+\mu_3(\chi_L^\dag\Phi\chi_R+\chi_R^\dag\Phi^\dag\chi_L).~~~~~~
\label{Scalar Potential}
\eea
We follow the theorems reported earlier ~\cite{PING1993109,Kannike:2012pe}, to ensure that
the matrix of the quartic terms, which are dominant at higher values of the
fields, is copositive. The ref.~\cite{Ashry:2013loa} presents a detailed study
on the conditions which keep the potential Eq.(\ref{Scalar Potential})
bounded from below.
After symmetry breaking there are ten scalars that remain as physical Higgs
bosons in ALRM, of which, four are charged Higgs bosons ($M_{H_1^\pm}, M_{H_2^\pm}$),
two are pseudo-scalar Higgs
bosons ($M_{A_1}, M_{A_2}$), and the remaining four are CP-even neutral Higgs
bosons ($M_{H}, M_{H_1}, M_{H_2}, M_{H_3}$).
The lightest neutral eigenstate $H$ is the SM-like Higgs, whose
mass is fixed to be $125.09$ GeV.
For charged Higgs $H_1^\pm$, the diagonalizable matrix is related to angle $\beta$ with $\tan\beta=k/v_L$.
However, for $H_2^\pm$, the diagonalizable matrix is related to angle $\zeta$ with $\tan\zeta=k/v_R$.
The vevs of $v_R$ is much larger than $v_L$. From this point, coupling of $H_1^\pm$ to the SM particles is stronger than that with $H_2^\pm$.
The LEP experiments have been conducted to
search for charged bosons via pair charged Higgs production.
The data statistically combined by four experiments (ALEPH, DELPHI, L3 and OPAL)~\cite{Abbiendi:2013hk}
showed that the mass of charged Higgs boson must be greater than 80 GeV.
This lower limit will be used as a reference in our calculations.
In recent past, ATLAS and CMS have also searched for the charged boson masses ranging from 200 to 2000 GeV~\cite{Atlas:2016se,CMS:2017se},
and constraints for some models such as hMSSM are given.

\section{NUMERICAL RESULTS AND DISCUSSION}

\subsection{ALRM effects in Higgs decay}

Each channel of discovered Higgs has already been detected by
CMS and ATLAS experiment groups. The decay signal strengths of Higgs to $ZZ$
bosons and $b\bar{b}$ pair are given in Table~\ref{strength},
independently by CMS~\cite{Sirunyan:2017exp,Chatrchyan:2013zna}, ATLAS~\cite{Aaboud:2017vzb,Aaboud:2017xsd}.

\begin{table}[H]
\caption{The decay signal strengths of Higgs to ZZ
bosons and $b\bar{b}$ pair given by CMS and ATLAS collaborations. $\mu_{XX}$ stands for $\mu(H \rightarrow XX)$.}
\label{strength}
\begin{center}
\begin{tabular}{l|crc}
\hline
\hline
                    & CMS & ATLAS    \\\hline
$\mu_{ZZ}$          &  $1.05^{+0.19}_{-0.17}$ & $1.29^{+0.20}_{-0.18} $ \\\hline
$\mu_{b\bar{b}}$    &$0.81^{+0.45}_{-0.43}$  &$0.90 \pm 0.18$    \\
\hline
\hline
\end{tabular}
\end{center}
\end{table}
Due to the detectors' limits and the defects of hardron collider, the data in Table~\ref{strength} may deviate from the real results, especially in $H\rightarrow b\bar{b}$.
The Higgs signal strength in a particular final state XX $\mu_{XX}$ is defined as
\bea
\mu_{XX} &=& \frac{\sigma}{\sigma^{\text{SM}}} ~\frac{\text{BR}(H\to XX)}{\text{BR}(H\to XX)^{\text{SM}}} \nonumber\\
&=& \frac{\Gamma(H\to gg)}{\Gamma(H\to gg)^{\text{SM}}}
~\frac{\Gamma_{\text{tot}}^{\text{SM}}}{\Gamma_{\text{tot}}} ~\frac{\Gamma(H\to XX)}{\Gamma(H\to XX)^{\text{SM}}}\nn\\
&=& \kappa_{gg} \cdot \kappa_{\text{tot}}^{-1} \cdot \kappa_{XX}~,%
\eea
where $\sigma$ stands for the total Higgs production cross
section at LHC and $\text{BR}(H\to XX)$ is the corresponding branching ratio. The total decay width of Higgs can be considered as the sum of some dominant Higgs partial decay widths. In Eq.(\ref{Fcoup}), we can see that the Yukawa couplings $Y_{Ht\bar{t}}$ and $Y_{Hb\bar{b}}$ in ALRM may be changed by adding a factor of $\frac{T_{\Phi}}{\sin\beta}$ and $\frac{T_{L}}{\cos\beta}$, respectively, from the SM values.

\begin{figure}[H]
\onefigure[scale=0.8]{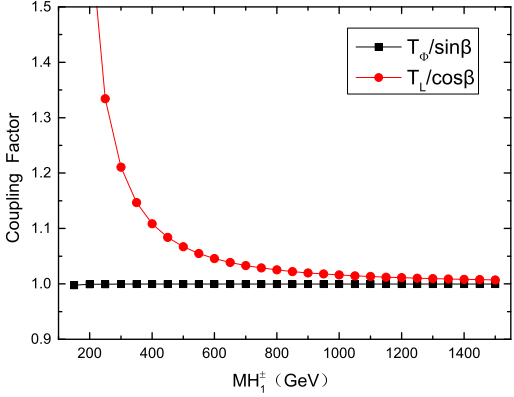}
\caption{The relation between $\frac{T_{\Phi}}{\sin\beta}$ and $\frac{T_{L}}{\cos\beta}$ as a function of $M_{H^\pm_{1}}$.}
\label{thl}
\end{figure}

In Fig.~\ref{thl}, the effect of these two factors are plotted to show that both
of them tend to be 1 while $\frac{T_{L}}{\cos\beta}$ of
$Y_{Hd\bar{d}}$ is greater than 1 particularly in the region of small $M_{H^\pm_{1}}$ with large $\tan\beta$.
$Y_{Hb\bar{b}}$ strongly depends on the mass of $H^\pm_{1}$.
However, the total decay width of Higgs boson remains very close to the SM result,
$\kappa_{\text{tot}}\simeq1$, when the mass of $H^\pm_{1}$ is big enough.

As for $\kappa_{gg}$, this channel is mainly propagated through the top quark triangle loop diagram and the extra quark $d'$ can be neglected due to the suppression of its coupling with SM-like Higgs. From Fig.\ref{thl}, we can see that the adding factor to top Yukawa coupling can be almost 1, making the top Yukawa coupling unchanged from the SM result. Therefore, the ratio $\kappa_{gg}=\Gamma(H\to~gg)/\Gamma(H\to~gg)^{\text{SM}}$ can be considered as 1.

Now, we turn to the SM-like Higgs decay into $ZZ$ in ALRM. For the kinematics forbidden, we compute $H \to ZZ$ via $H \to ZZ^* \to Zff$, where $f = e,\mu,\tau$ and $u,d,c,s,b$ quark.

It is worth mentioning that the parameters $\lm_3,~\al_{12}$ and $M_{H^\pm_{2}}$ are not
sensitive in the numerical results, only $0<\lm_3<\sqrt{4\pi}$ and $\al_{12}>0$, to be
consistent with the perturbative unitarity and the minimization and boundedness from
below conditions Eqs.(21--24) in ref.~\cite{Ashry:2013loa}.
The relevant input parameters are chosen as~\cite{Olive:2016xmw}:

\bea
\al(0) = 137.035999,~~m_W = 80.385 ~GeV,\nonumber \\
m_Z = 91.1876 ~GeV,~~m_H = 125.09~GeV,\nonumber \\
M_{H^\pm_{2}} = 150~GeV,~~~~M_{W'} = 4000~GeV,\nonumber \\
\al_1 = \al_2=1,~~~~~~~~~\lm_3 = 1.5~.
\eea

\begin{figure}[H]
\onefigure[scale=0.8]{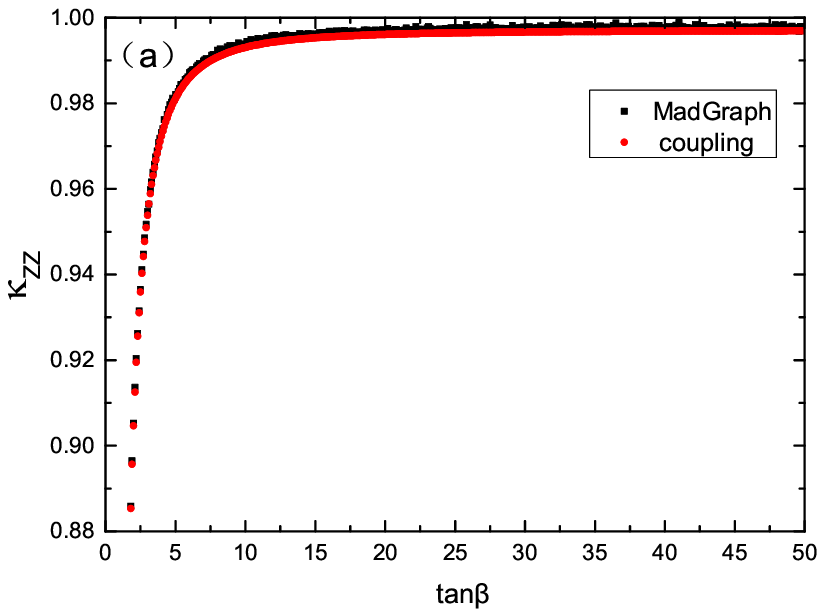}
\onefigure[scale=0.8]{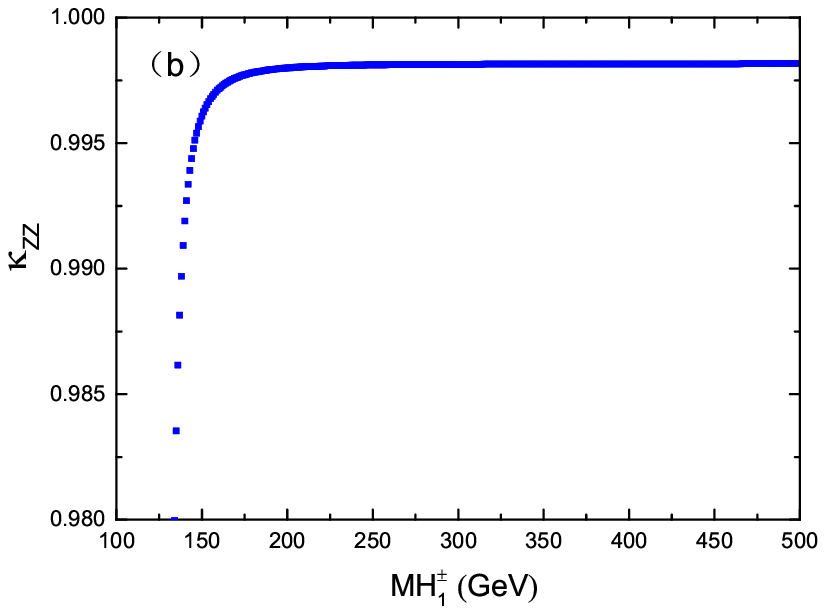}
\caption{$\kappa_{ZZ}$ as functions of $\tan\beta$ and $M_{H^\pm_{1}}$
 for the parameters $\lm_3,~\al_{12}$, and $M_{H^\pm_{2}}$, respectively.}
\label{kzz}
\end{figure}

We have used Feynrules~\cite{Alloul20142250,Degrande20121201} to generate the model files and MadGraph~\cite{Alwall2014}
to calculate the numerical values of the cross-sections.
In Fig.~\ref{kzz}, we display the results of $\kappa_{ZZ} =
\Gamma(H\to ZZ)/\Gamma(H\to ZZ)^{\text{SM}}$ as functions of $\tan \beta$ and $M_{H^\pm_{1}}$. This Figure confirms our theoretical expectation and shows that $\kappa_{ZZ}$ can slightly deviate from 1. In Fig.~\ref{kzz}(a) $\kappa_{ZZ}$
is calculated by MadGraph and the relevant couplings $\kappa_{ZZ} =
g(HZZ)_{\text{ALRM}}^2/g(HZZ)_{\text{SM}}^2$, respectively. Obviously, the results
are the same in both the cases.
In Fig.~\ref{kzz}(b), with increase in $M_{H^\pm_{1}}$, the value rapidly
increases and stabilizes when $M_{H^\pm_{1}}\gtrsim 200~GeV$.
Hence, in the following calculation $M_{H^\pm_{1}}$ is equal to 200 GeV, unless
otherwise stated. In this case, it is clear that the signal strength $\mu_{zz}$ is also close to the SM expectation and can be consistent with CMS experimental results.
It is remarkable that all signal strengths of Higgs decay channels in ALRM
are close to SM results with MadWidth~\cite{Alwall2015312} automatically
computing decay widths.

\begin{figure}[H]
\onefigure[scale=0.8]{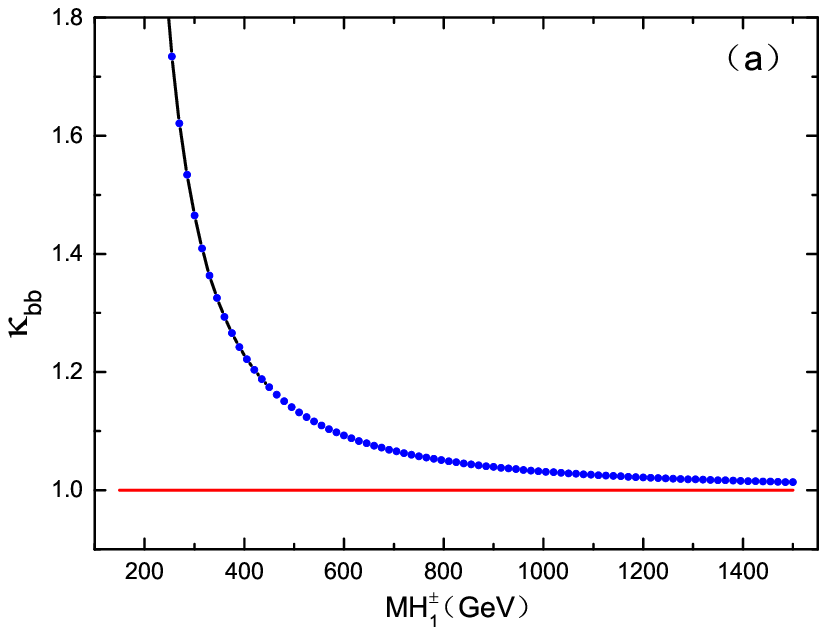}
\onefigure[scale=0.8]{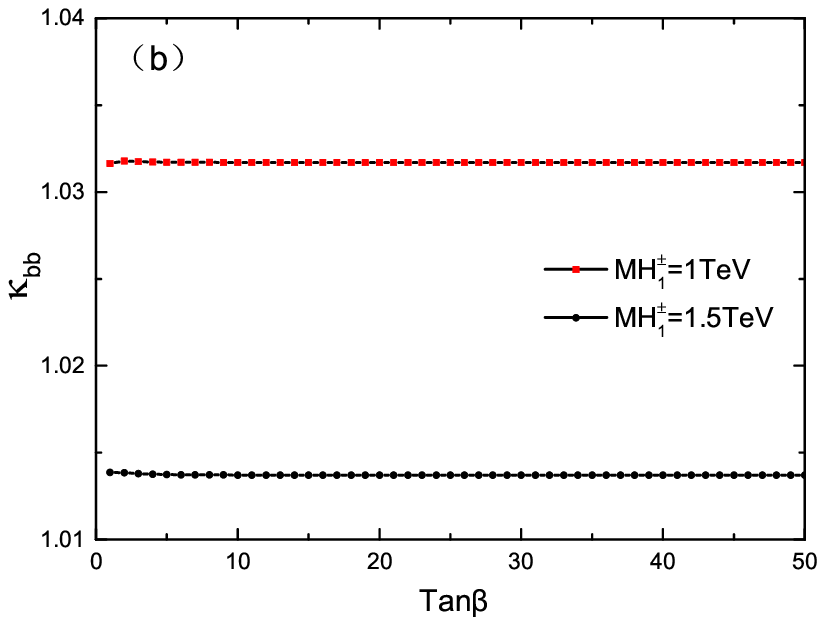}
\caption{(a)$\kappa_{b \bar{b}}$ as a function of $M_{H^\pm_{1}}$ for the
parameter $\tan \beta$=50; (b)$\kappa_{b \bar{b}}$ as a function of $\tan \beta$ for the
parameter $M_{H^\pm_{1}}=600$ GeV.}
\label{hbb}
\end{figure}

In Fig.~\ref{hbb}(a), $\kappa_{b \bar{b}}$ is plotted as a function of $M_{H^\pm_{1}}$. This figure shows that the decay channel $H\to b\bar{b}$ is more sensitive to $M_{H^\pm_{1}}$ than the channel $H\to ZZ$.
In addition, it is remarkable that the decay width of this channel in ALRM is slightly larger than that in SM. In Fig.~\ref{hbb}(b), $\kappa_{b \bar{b}}$ is plotted as a function of $\tan \beta$, it is clearly seen that $\tan \beta$ has negligible impact on $\kappa_{b \bar{b}}$. From Fig.~\ref{hbb}, we find that the decay channel $H \to b\bar{b}$ is more sensitive to $M_{H^\pm_{1}}$ than to $\tan \beta$. And $\kappa_{b \bar{b}}$ decreases significantly with increasing $M_{H^\pm_{1}}$. Cause for the constraints on $M_{H^\pm_{1}}$ from the decay channel $H \to b \bar{b}$, $M_{H^\pm_{2}}$ can be varied in a larger parameter space than $M_{H^\pm_{1}}$.

\subsection{ALRM effects in $e^+ e^- \to HZ$}

\begin{figure}[H]
\onefigure[scale=0.55]{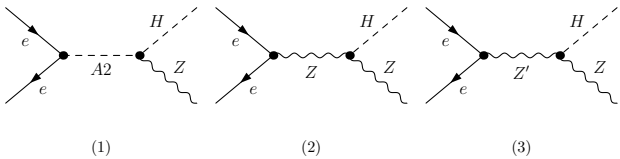}
\caption{Feynman diagrams for the process $e^+ e^- \to HZ$ with a
new scalar $A_2$, a heavy boson $Z'$ propagator and some new couplings.}
\label{eehz}
\end{figure}

In this subsection the production of $e^+ e^- \to HZ$ at CEPC and future $e^+~e^-$ colliders is presented and Feynman diagrams of this process
are displayed in Fig.~\ref{eehz}. As the contributions from t channel are negligibly small, we only give the Feynman diagrams in s channel. As can be seen from Fig.~\ref{eehz}, a new scalar $A_2$ and a heavy boson $Z'$ are added in the propagators. The relevant input parameters are chosen as described above.

\begin{figure}[H]
\onefigure[scale=0.8]{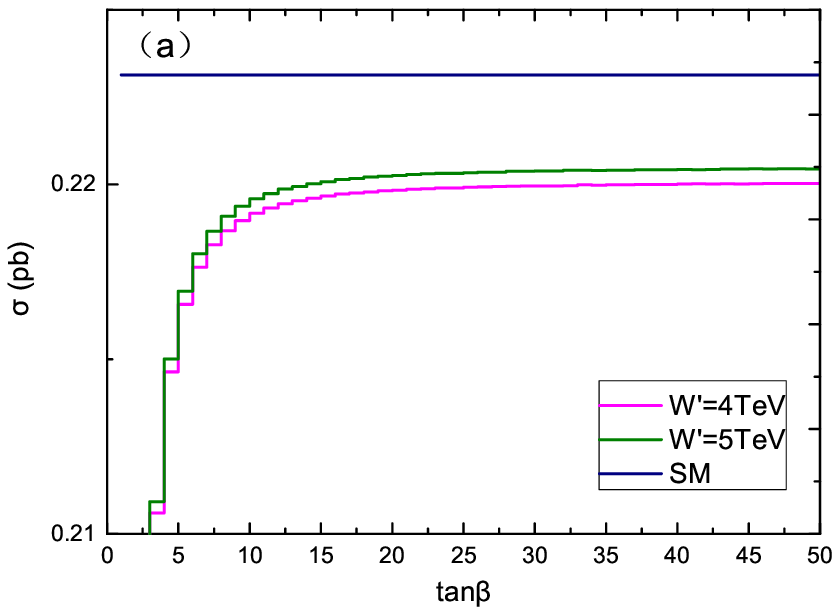}
\onefigure[scale=0.8]{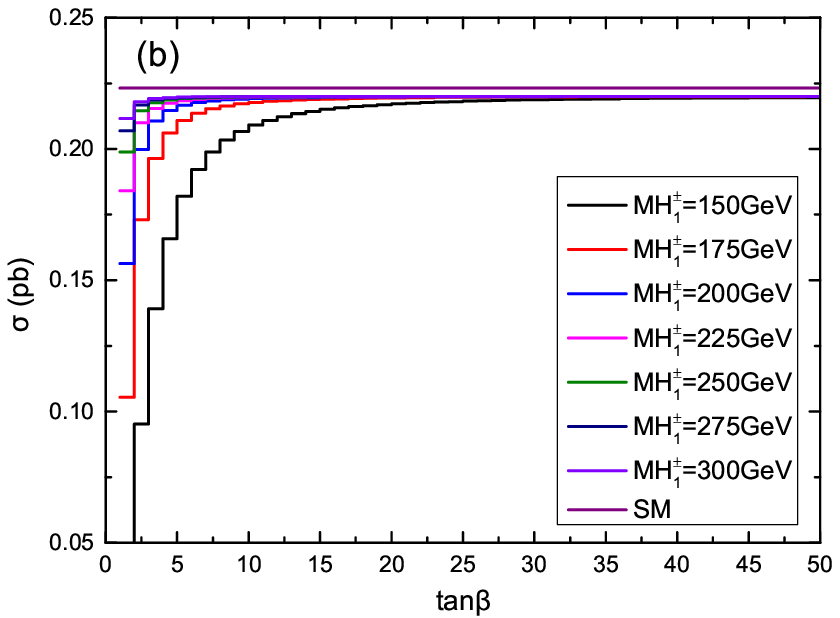}
\caption{Cross-section as a function of $\tan\beta$ for the
parameters $M_{W'}$ and $M_{H^\pm_{1}}$ at $\sqrt{s}=240$ GeV.}
\label{cs-tb-240}
\end{figure}

The cross section as a function of $\tan\beta$ with the mass of heavy boson $W'$ varying from 4 TeV to 5 TeV and charged boson $H^\pm_{1}$ varying from 150 GeV to 300 GeV is shown in Fig.~\ref{cs-tb-240} at $\sqrt{s}=240$ GeV. Obviously, the total cross-sections are all less than those in SM from Fig.~\ref{cs-tb-240}. From Ref.~\cite{Ashry:2013loa}, the mass of $A_2$ become small when $\tan\beta$ and $M_{H^\pm_{1}}$ are small, and $M_{W'}$ influences the heavy boson $Z'$. In Fig.~\ref{cs-tb-240}, due to the large $M_{W'}$, the contribution from $Z'$ propagator is small and the contribution from $A_2$ is mainly in small $\tan\beta$ region. Fig.~\ref{cs-tb-240}(b) shows the discrepancies from $A_2$ propagator become larger with smaller $M_{H^\pm_{1}}$. At the other two collision energies, the same tendency can be obtained which we didn not show.

The total cross-sections and corresponding relative ALRM discrepancies are tabulated in Table~\ref{discrepancy} with $\tan\beta=50$, $M_{H^\pm_{1}}=200$ GeV and $M_{W'}=4$ TeV at $\sqrt{s}=240$ GeV, 500 GeV, and 1 TeV, respectively. The relative ALRM discrepancy is defined as $\delta $=$ (\sigma_{ALRM}-\sigma_{SM})/\sigma_{SM}$. In this table, we can see the relative ALRM discrepancies are increasing with $\sqrt{s}$. It is worth mentioning that the Higgs sector in ALRM is very similar to that in 2HDM, where one Higgs doublet couples to up-quarks and the second couples to down-quarks. Therefore, it does not lead to any flavor changing neutral current problem and light charged Higgs is phenomenologically acceptable.

\begin{table}[H]
\caption{The total cross-sections and corresponding relative ALRM discrepancies for the process $e^+ e^- \to HZ$ at $\sqrt{s}=240$ GeV, 500 GeV and 1 TeV respectively.}
\label{discrepancy}
\begin{center}
\begin{tabular}{l|crc}
\hline
\hline
$\sqrt{s}$(GeV)   & ~$\sigma_{\text{ALRM}}$(fb) & ~$\sigma_{\text{SM}}$(fb)  & ~$\delta$(\%)$$  \\
[0.5ex]\hline
240               & ~$220.06$                   & ~~$223.1$         & ~~~$-1.36$  \\
[0.5ex]\hline
500               & ~$51.66$                    & ~~$53.22$         & ~~~$-2.93$  \\
[0.5ex]\hline
1000              & ~$11.13$                    & ~~$11.94$         & ~~~$-6.78$  \\
\hline
\hline
\end{tabular}
\end{center}
\end{table}

\begin{figure}[H]
\onefigure[scale=0.8]{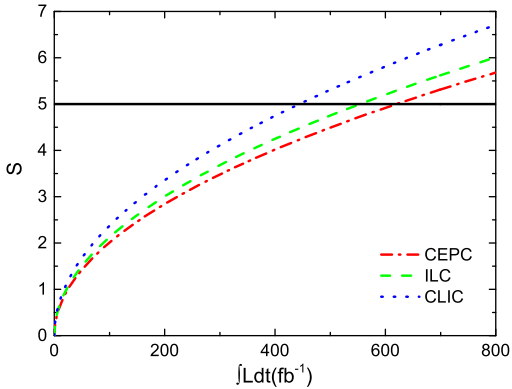}
\caption{Discovery significance determined by $\frac{n_S}{\sqrt{n_{tot}}}$ as a function of integrated
luminosity for CEPC, ILC and CLIC at $\sqrt{s}=240$ GeV, 500 GeV and 1 TeV respectively
.}
\label{lumi}
\end{figure}

We next analyze the discovery significance, which is calculated using the formula
$\frac{n_S}{\sqrt{n_{tot}}}$, where
$n_S=\int Ldt\times(\sigma_{ALRM}-\sigma_{SM})$ is the number of discrepancy events
and $n_{tot}=\int Ldt \times\sigma_{SM}$ is the number of total events.
The plot of discovery significance as a function of integrated luminosity for CEPC, ILC, and CLIC is depicted in Fig.~\ref{lumi}. From Fig.~\ref{lumi}, one can see that the integrated luminosity of all the three colliders can reach several hundreds, specifically at $619.72~fb^{-1}$(CEPC), $553.80~fb^{-1}$(ICL),
and $443.93~fb^{-1}$(CLIC) when the discovery significance is $5\sigma$.
By contrast, CLIC seems to have an advantage in detecting it. It is worth mentioning that the discovery significance shown in Fig.~\ref{lumi} is calculated with no kinetic cuts. It cannot be treated as a serious
result.

\subsection{ALRM effects in $e^+ e^- \to HZ \to l^+~l^-~ b ~\bar{b}$}

\begin{figure}[H]
\onefigure[scale=0.55]{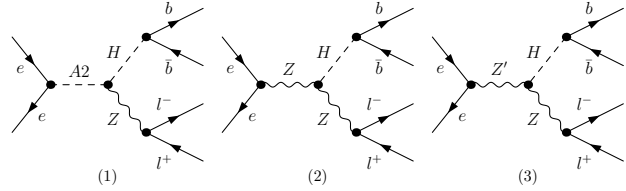}
\caption{Feynman diagrams for subsequent decay of the SM-like Higgs boson into a pair of bottom quarks and Z boson into a leptonic pair, where $l = e, ~\mu~and~\tau$, associated with ALRM.}
\label{eehzdecay}
\end{figure}

In this subsection we will analyze and compute the cross-section for production and subsequent decay at $\sqrt{s}=240$ GeV, 500 GeV and 1 TeV, respectively. Feynman diagrams for subsequent decay of the SM-like Higgs boson into a pair of bottom quarks and Z boson into opposite-sign dilepton, where $l = e,~\mu$ and $\tau$, are shown in Fig.~\ref{eehzdecay} associated with ALRM. When doing the numerical calculation, the mass of fermions is chosen as follows£º

\bea
M_{e}=5.11 \times 10^{-4}~GeV,~M_{\mu}=1.0566 \times 10^{-1}~GeV,\nonumber\\
M_{\tau}=1.777~GeV,~~~~~~~~M_{b}=4.7~GeV.~~~~~~~~~~~~~~
\label{lmass}
\eea

In dealing with the sequential Z boson leptonic decay and Higgs decay,
the naive narrow-width approximation(NWA) method is used to acquire the total cross-section.
Hence, cross-section for this process can be approximately written as:

\bea
&&\sigma\left(e^+ e^- \rightarrow l^+
l^- b\bar{b}\right)\simeq \sigma\left(e^+e^-
\rightarrow HZ\right)\nn
\\&&~~~~~~\times\text{BR}\left(H\rightarrow b\bar{b}\right)
\times\text{BR}\left(Z \rightarrow l^+ l^-\right).
\eea

In order to get the precise branch ratio of $H \to b\bar{b}$ in ALRM, we set the K-factor of each channel the same as that in SM.
By adopting HDECAY program~\cite{Djouadi:1997yw}, dominant Higgs partial decay widths are computed that are tabulated in Table~\ref{hdecay}. Here, the scale $\mu$ of the Yukawa coupling $y_Q(\mu)=m_Q(\mu)/v$ is used as the mass of Higgs in the calculation, while $m_Q(\mu)$ is the running mass of
heavy quark. The total cross-section $\Gamma_{tot}=\Gamma_{b\bar{b}}~+~\Gamma_{c\bar{c}}~+~\Gamma_{\tau^+\tau^-}~+~\Gamma_{ZZ}~+~
\Gamma_{WW}~+~\Gamma_{gg}
=3.93\times10^{-3}$~GeV. Obviously, the branch ratio of $H \to b\bar{b}$ is $58.27\%$ in SM and experiment shows that SM prediction for the decay branching fraction of Higgs boson with mass around 125.09 GeV to $b\bar{b}$ is $57.5\%$~\cite{Khachatryan:2016vau}.
The leading order(LO) results in SM computed by MadWidth and the corresponding
K-factors~($\frac{\sigma_{\text{HDECAY}}}{\sigma_{\text{MadWidth}}}$) are also included
in Table~\ref{hdecay}. The decay width of $H \to gg$ is directly used
result in SM by HDECAY. In this manner, we can estimate the branch ratio of
$H \to b\bar{b}$ in ALRM as a function of $M_{H^\pm_{1}}$. As for the branch ratio of
$Z \to l^+l^-$, the result is $10.31\%$ in both SM and ALRM which is independent of $M_{H^\pm_{1}}$.

\begin{table}[H]
\caption{Higgs boson partial decay widths in the framework of SM
computed by HDECAY and MadWidth, and the corresponding K factors.}
\label{hdecay}
\begin{center}
\begin{tabular}[scale=0.8]{l|crc}
\hline
\hline
Decay mode           & HDECAY  & MadWidth  & K factor  \\
                     & [MeV] &[MeV]~~~ &   \\
\hline
$H\to b\bar{b}$         & $2.29$        & $1.8320~~~$      & ~~~$1.2514$  \\
\hline
$H\to c\bar{c}$         & $0.1043$      & $0.0824~~~$     & ~~~$1.2658$   \\
\hline
$H\to \tau^+\tau^-$     & $0.2474$       & $0.2499~~~$     & ~~~$0.9903$   \\
\hline
$H\to ZZ^* $     & $0.1047$        & $0.0607~~~$      & ~~~$1.7249$   \\
$~~~\to Zff $     &&&                                             \\
\hline
$H\to WW^*  $      &$0.8561$     & $0.4863~~~$      & $1.7604$   \\
$~~~\to Wff  $      &   &       &  \\
\hline
$H\to gg $               &$0.3261$      &                           &   \\
\hline
\hline
\end{tabular}
\end{center}
\end{table}

The total cross-sections and corresponding relative discrepancies are shown in Table~\ref{nwa} at $\sqrt{s}$ = 240 GeV,~500 GeV and 1~TeV respectively, where the relative deviation is defined as $\delta $=$ (\sigma_{ALRM}-\sigma_{SM})/\sigma_{SM}$. From Table~\ref{nwa} one finds that with the increase of $\sqrt{s}$, the cross-sections in both ALRM and SM decrease. While the corresponding relative discrepancies are increasing significantly, which will be phenomenologically accessible.

\begin{figure}[H]
\onefigure[scale=0.8]{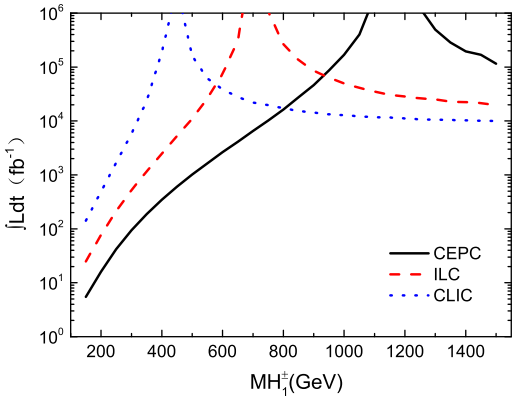}
\caption{The integrated luminosity needed for $5\sigma$ discovery
significance as a function of $M_{H^\pm_{1}}$ for CEPC, ILC, and CLIC at
$\sqrt{s}=240$~GeV,~500~GeV~and~1~TeV, determined by $\frac{n_S}{\sqrt{n_{tot}}}$.}
\label{sigma}
\end{figure}

\begin{table}[H]
\caption{ The total cross-sections and corresponding relative discrepancies
for the process $e^+ e^- \to HZ \to l^+~l^-~ b ~\bar{b}$ by NWA with $M_{H^\pm_{1}}=600~GeV$
at $\sqrt{s}=240$ GeV, 500 GeV and 1 TeV respectively.}
\label{nwa}
\begin{center}
\begin{tabular}{l|crc}
\hline
\hline
$\sqrt{s}$(GeV)   & $\sigma_{\text{ALRM}}$(fb) & $\sigma_{\text{SM}}$(fb)  & $\delta$(\%)$$  \\
[0.5ex]\hline
240               & ~$13.522$                    & ~~$13.397$            & ~~~$0.93$  \\
[0.5ex]\hline
500               & ~$3.174$                  & ~~$3.197$                & ~~~$-0.72$   \\
[0.5ex]\hline
1000              & ~$0.684$                 & ~~$0.717$                & ~~~$-4.60$   \\
\hline
\hline
\end{tabular}
\end{center}
\end{table}

To analyze the feasibility of experiment, the discovery significance is chosen to be the same as last subsection:$\frac{n_S}{\sqrt{n_{tot}}}$, where
$n_S=\int Ldt\times BR(Z\to l^+l^-)\times (\sigma_{ALRM}\times BR_{ALRM}(H\to b\bar{b})-\sigma_{SM}\times BR_{SM}(H\to b\bar{b}))$ is the number of discrepancy events and $n_{tot}=\int Ldt\times BR(Z\to l^+l^-)\times \sigma_{SM}\times BR_{SM}(H\to b\bar{b})$ is the total number of events. In Fig.~\ref{sigma}, we depict the integrated luminosity needed for $5\sigma$ discovery significance as a function of $M_{H^\pm_{1}}$ for CEPC, ILC and CLIC determined by $\frac{n_S}{\sqrt{n_{tot}}}$. For $e^+ e^- \to HZ \to l^+~l^-~ b ~\bar{b}$ process,
the discrepancies between SM and ALRM are mainly influenced by the cross-section of $e^+e^-\to HZ$ and the branching ratio of Higgs to $b\bar{b}$. The cross-sections of $e^+e^-\to HZ$ in ALRM is a little smaller than that in SM but the branching ratio of $H\to b\bar b$ is opposite. In the middle of $M_{H_1^\pm}$ region, two parts of the contribution counteract each other while the cross-section of $e^+ e^- \to HZ \to l^+~l^-~ b ~\bar{b}$ in ALRM and SM approach each other. Hence, in order to detect the discrepancies, the required integrated luminosity needs to be very large, which means that it is difficult to search new physics in the region. This corresponds to the peak in Fig.~\ref{sigma}.

\section{Summary}

In the present paper, we have analyzed Higgs decay in each channel, Higgs and Z boson associated production and decay at CEPC and future $e^+~e^-$ colliders in ALRM, motivated by superstring inspired $E_6$ model. We found that the contribution of charged Higgs boson $M_{H^\pm_{2}}$ to Higgs decay is negligible due to the large $\tan\beta$, while $M_{H^\pm_{1}}$ plays an essential role in the decay channel of $H \to b\bar{b}$ due to the mixing of scalars. In addition, the model predicts the signal strengths of Higgs decay, of $H \to ZZ$ in particular, that are consistent with SM expectations. We also analyzed the discrepancies of cross-sections about Higgs and Z boson production between ALRM and SM and found that it can be enhanced to 6.78$\%$ when $\sqrt{s}$ is increased to 1 TeV. Finally, we studied the sequential decay of Higgs and Z boson to $b\bar{b}$ and $l^+~l^-$, respectively, where $l=e, ~\mu$, and $\tau$, with NWA method. We found that the cross-sections of sequential decay are significantly dependent on the branch ratio of $H\to b\bar{b}$. We have also shown that the typical values of cross-sections are of ${\cal O}(1)$ fb which can be measured using future colliders.

\end{multicols}

\vspace{5mm}

\begin{multicols}{2}

\end{multicols}

\clearpage

%\end{CJK*}
\end{document}